\documentclass[%
 reprint,
superscriptaddress,
 amsmath,amssymb,
 aps,
]{revtex4-1}
\usepackage{ragged2e} 
\usepackage{booktabs}
\usepackage{multirow}

\usepackage{subcaption}

\usepackage{float}
\usepackage{caption}
\usepackage{bm} 
\usepackage{graphicx}
\usepackage{dcolumn}
\usepackage{bm}
\usepackage{hyperref}

\begin{document}

\preprint{APS/123-QED}

\title{Probing black holes via quasinormal modes in a dark energy-induced dark matter}
\author{Jie Liang}
\affiliation{%
 College of Physics,Guizhou University,Guiyang,550025,China
}%

\author{Dong Liu}
\affiliation{Department of Physics, Guizhou Minzu University, Guiyang, 550025, China}

\author{Hao-Jie Lin}
\affiliation{
    Key Laboratory of Cosmology and Astrophysics (Liaoning), \\
    College of Sciences, Northeastern University, Shenyang 110819, China
}
\affiliation{
    Institute for Theoretical Physics and Cosmology, \\
    Zhejiang University of Technology, Hangzhou, 310032, China
}
\affiliation{
    United Center for Gravitational Wave Physics (UCGWP), \\
    Zhejiang University of Technology, Hangzhou, 310032, China
}

\author{Zheng-Wen Long}%
\email{zwlong@gzu.edu.cn (corresponding author)}
\affiliation{%
 College of Physics,Guizhou University,Guiyang,550025,China
}%


\begin{abstract}

This study delves into the existence of dark matter around supermassive black holes in galactic cores using a novel gravitational model. By analyzing gravitational waves emitted during the ringdown phase of black holes under different field perturbations, we explore the potential for detecting dark matter. The model hypothesizes that the dark matter distribution around black hole is driven by a mechanism where dark energy endows gravitons with mass, thereby forming a new spacetime structure. Results reveal that as relevant parameters increase, the quasinormal modes (QNMs) exhibit a gradual reduction in real parts, with negative imaginary parts whose absolute values also decrease. Moreover, compared to gravitational wave signals from Schwarzschild black hole without dark matter, this system demonstrates significant differences in oscillation modes and frequencies. This achievement not only validates the self-consistency of the new gravitational model but also lays a theoretical foundation for subsequent gravitational wave detection within dark matter. Simultaneously, it provides new theoretical support for understanding the mechanism of dark energy in large-scale cosmic structures and broadens the research perspective on the relationships between black hole physics, dark matter, and dark energy.

\end{abstract}

\maketitle


\section{\label{sec:level1}Introduction}
Since the LIGO and Virgo successfully detected the gravitational wave signals generated by the merger of binary black holes in 2016 \cite{LIGOScientific:2016aoc,LIGOScientific:2016sjg,LIGOScientific:2017vwq}, the existence of gravitational waves has been   confirmed. This significant discovery has sparked a new upsurge in observing gravitational waves \cite{KAGRA:2023pio,LIGOScientific:2021qlt,LIGOScientific:2020stg} and has promoted significant progress in the field of astronomy. During the coalescence process of binary black holes, the system usually undergoes the initial inspiral, merger, and ringdown stages \cite{LIGOScientific:2017ycc}. The ringdown stage can be characterized by the superposition of complex frequency damped exponentials and is defined as QNMs \cite{Chandrasekhar:1984siy,Maggiore:2019uih,Konoplya:2011qq,Vishveshwara:1970zz}. Specifically, the real part represents the oscillation frequency of the gravitational wave (i.e., the vibration period), while the imaginary part represents the damping rate of the vibration, reflecting the rate at which the vibration weakens over time \cite{Kokkotas:1999bd,Cardoso:2019rvt}. In the context of black holes, QNMs can describe the process of emitting gravitational waves after being perturbed. It is worth noting that these frequencies depend only on the parameters identifying the black holes. Therefore, QNMs can serve as a crucial tool for us to explore black holes \cite{Rahman:2021kwb,Liu:2022ygf,Liu:2023vno}. Through the analysis of these QNMs, researchers can obtain detailed information about the structure of black holes, such as the mass and spin of black holes, and may even reveal the extreme material environment surrounding black holes \cite{Liu:2021xfb, Yang:2022ifo, Liu:2024xcd}. 

Dark matter, as an important component in the universe, has a significant impact on the astronomical environment around black holes \cite{Ferrer:2017xwm, Nampalliwar:2021tyz, Xu:2021dkv}. Research shows that the contribution rate of dark matter to the total mass of galaxies can be as high as $90\%$. Meanwhile, astronomical observations and theoretical predictions indicate that supermassive black holes exist at the centers of most galaxies. The powerful gravitational fields of these black holes will significantly affect the distribution patterns of the surrounding dark matter, causing the dark matter to exhibit specific structures in the regions close to the black holes \cite{EventHorizonTelescope:2019dse,EventHorizonTelescope:2019ggy,Navarro:1995iw}. Gondolo et al. adopted the Newtonian approximation method and derived for the first time the density distribution of dark matter around the black hole at the center of the Milky Way \cite{Gondolo:1999ef}, and pointed out that under the accretion of the black hole, dark matter will form a spike structure. Xu et al. combined the NFW profile of the dark matter halo with the spike structure and further derived the space-time geometry of black holes immersed in dark matter \cite{ Xu:2020jpv}. Liu et al. utilized the mass model of M87* and the Einasto profile in the cold dark matter model to describe the structure of the dark matter halo \cite{ Liu:2023oab}.
However, to date, researchers have primarily employed numerical simulation methods to investigate the effects of dark matter, while its existence is inferred from observations of gravitational effects on galaxies and other large-scale cosmic structures \cite{Navarro:1995iw,deBlok:2009sp,Li:2012zx,Haroon:2018ryd,Konoplya:2019sns,Jusufi:2019nrn,Narzilloev:2020qtd,Caputo:2023cpv}. Presently, the nature of this enigmatic substance remains elusive \cite{Gonzalez:2023rsd}. On the other hand, with the continuous deepening of astronomical observations, especially the precise measurements of supernova explosions and cosmic microwave background radiation, the results show that the universe is in a stage of accelerated expansion, and this phenomenon reveals the possibility of the existence of dark energy \cite{SupernovaCosmologyProject:1998vns,SupernovaSearchTeam:1998fmf,SupernovaSearchTeam:1998cav, Inan:2022idt,Beck:2005pr}. The properties of dark energy are usually considered to be related to the quantum vacuum energy and may eventually lead the universe into a de Sitter state of permanent expansion \cite{Carroll:2000fy, Caravano:2024tlp}. However, if dark energy is the cause of the quantum vacuum energy, the huge difference between the theoretically calculated vacuum energy density and the observed value has sparked disputes about the nature of dark energy. Recently, Nader Inan et al., inspired by the Higgs mechanism that gives photons mass in superconductors, proposed that dark energy may give gravitons mass \cite{Inan:2024noy}. This causes the classical Newtonian gravitational potential to be corrected into a form similar to the Yukawa potential \cite{Inan:2024noy}. Using the Yukawa potential, the dark matter model in cosmology can be effectively obtained  \cite{Jusufi:2024ifp}. Moreover, it can connect baryonic matter, emergent dark matter and dark energy \cite{Gonzalez:2023rsd, Jusufi:2023xoa}.

Black holes serve as a large laboratory for testing gravitational theories. When this model is combined with black holes, a unique spacetime geometry is formed \cite{Pantig:2024rmr}. Against this backdrop, it has become an interesting topic to conduct in-depth research on the behavior of the quasi-normal modes of this spacetime geometry under the perturbations of various fields. By calculating and analyzing the quasi-normal modes under this specific spacetime geometry, we can not only test the validity of this emerging dark matter model, but also are expected to reveal the potential interaction mechanism between dark matter and black holes, thus providing new clues for an in-depth understanding of the nature of dark energy and the behavior of gravity in strong gravitational fields.

The structure of this paper is arranged as follows: In Section \ref{sec:2}, we introduce the black hole metric of dark matter formed by dark energy \cite{Pantig:2024rmr}, and constrain the ranges of the parameters $\lambda_{G}$ and $\eta$ by simulating the $M87^*$ galaxy. The equations of motion under the perturbations of the matter field and gravitational perturbations are derived. In Section \ref{sec:3}, we introduce two numerical calculation methods commonly used in the study of the quasinormal mode problem, namely the WKB approximation method and the time-domain method. In Section \ref{sec:4}, the ringing waveform is obtained by solving in the time domain, and the variation of the quasinormal mode frequencies of black holes is explored. The last section is our summary. In this paper, we mainly use the units of $(c = G = 1)$.

\section{\label{sec:2}Perturbations of the black - hole model field of the dark - matter under the domination of dark energy}
In this section, we study the black hole model of dark matter dominated by dark energy, which adopts the spherically symmetric metric derived from the literature. The specific form is as follows \cite{Pantig:2024rmr}:

\begin{equation}
ds^2=-f(r)dt^{2}+\frac{1}{f(r)}dr^{2}+r^{2}d\Omega^{2}
\end{equation}
where, \(d\Omega^{2} = d\theta^{2} + \sin^{2} \theta\, d\phi^{2}\), representing the angular element metric of the spherically symmetric space. \(f(r)\) is the metric function that describes the spacetime outside the black hole and is given by:

\begin{equation}
f(r)=e^{-\frac{2\eta}{r}\,e^{-\frac{r}{\lambda_G}}}-\frac{2M}{r}
\end{equation}
In this metric, \(\lambda_{G}\) represents a Newtonian gravitational shielding length parameter with a superconducting-like property arising from the existence of dark energy. This parameter describes the shielding effect that gravity may experience on a large scale, that is, the gravitational interaction may be weakened or shielded after exceeding a specific length scale \(\lambda_{G}\). Specifically, \(\lambda_{G}\) is defined as:
\(\lambda_{G}=\frac{1}{\sqrt{2\Lambda}}\), where \(\Lambda\) is a cosmological constant that is closely related to the properties of dark energy. Another parameter \(\eta\) characterizes the distribution of dark matter and is related to the mass distribution of baryons. When \(\eta = 0\), the above metric degenerates into the classic Schwarzschild Black Hole metric. 

\begin{figure}[htbp]
    \centering 
    \includegraphics[width=0.45\textwidth]{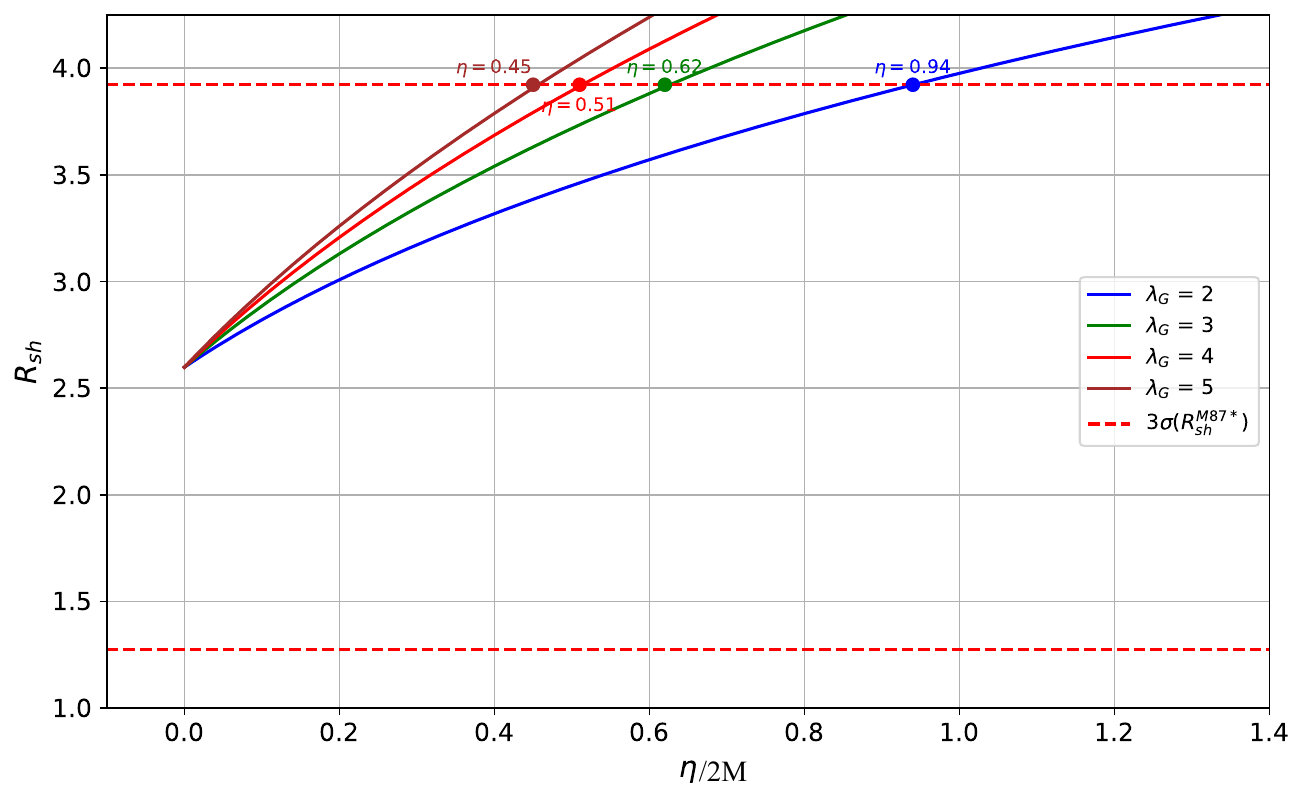} 
    \captionsetup{justification=raggedright,singlelinecheck=false} 
    \caption{Constraints on $\eta$ using the actual data of the Event Horizon Telescope (EHT). The red dashed line is the $3\sigma$ confidence interval of $M87^*$. The labeled value is the upper limit of $\eta$, and when $\eta = 0$, it represents the situation of a Schwarzschild black hole without being surrounded by dark matter.} 
    \label{fig:55} 
\end{figure}

Herein, it is of great significance to determine the parameter ranges. Our aim is to delimit the value ranges of the parameters $\eta$ and $\lambda_{G}$ at the $3\sigma$ (99.7\%) confidence level. The symbol $\sigma$ represents the probability that the true parameter values are contained within this interval at the preset confidence level. Considering the supermassive black hole in the M87* galaxy, at the $3\sigma$ confidence level, as reported in the literature \cite{Pantig:2024rmr}, the range of the black hole shadow radius $R_{sh}$ is $2.546M\leq R_{sh}\leq7.846M$. We re-analyze the variation of the shadow radius with respect to the parameters $\lambda_{G}$ and $\eta$ under the condition that the mass $M$ of the M87* black hole is $\frac{1}{2}$. The results are shown in Figure \ref{fig:55}. It can be observed that as $\lambda_{G}$ and $\eta$ increase, the black hole shadow radius $R_{sh}$ shows an obvious trend of change. Specifically, when $\lambda_{G}$ increases from a relatively small value, the shadow radius $R_{sh}$ also increases accordingly, which indicates that the gravitational shielding effect becomes more pronounced as $\lambda_{G}$ increases. Similarly, an increase in $\eta$ also leads to an increase in the shadow radius, reflecting the impact of the distribution of dark matter on the black hole shadow region. The red dashed line in the figure represents the $3\sigma$ confidence interval. 

Inspired by Figure \ref{fig:55} and literature \cite{Pantig:2024rmr}, we can finally constrain the ranges of the parameters \(\eta\) and \(\lambda_G\) as shown in Table \ref{table:1}.

\begin{table}[H]
\centering
\begin{tabular}{ccccc}
\hline
\multicolumn{1}{c}{BHs} & \multicolumn{1}{c}{Observatory} & \multicolumn{1}{c}{Fixed parameter} & \multicolumn{1}{c}{Range}  \\

\hline
\multirow{2}{*}{M87*} & \multirow{2}{*}{EHT} & $\eta = 0.4 $ & $0.09978 \leq \lambda_G \leq 6.113$ \ \\
& & $\lambda_G = 2$ &  $0 \leq \eta \leq 0.94 $ \\
\hline
\end{tabular}
\vspace{0pt} 
\caption{Upon the fixation of the values of parameters \(\eta\) and \(\lambda_G\), the observational data pertaining to the black hole shadow of M87*, as procured by the EHT, is harnessed to delimit the value range of an alternative parameter within the precincts of the \(3\sigma\) range. } 
\label{table:1}
\end{table}

\subsection{Perturbations of the Scalar Field and Electromagnetic Field}

In this section, we discuss the evolution equations of the massless scalar field and the electromagnetic field in the background of the spherically symmetric metric. The evolution equations of these fields are described in the generally covariant form as follows:
\begin{equation}
\frac{1}{\sqrt{-g}}\partial_{\mu}(\sqrt{-g}g^{\mu\nu}\partial_{\nu}\Phi) = 0,\label{eq:3}
\end{equation}
\begin{equation}
\frac{1}{\sqrt{-g}}\partial_{\mu}(F_{\rho\sigma}g^{\rho\nu}g^{\sigma\nu}\sqrt{-g}) = 0.\label{eq:4}
\end{equation}
Here, $g$ is the determinant of the metric tensor, $g^{\mu\nu}$ is the contravariant form of the metric, $\Phi$ is the wave function $\Phi(t,r,\theta,\phi)$ of the scalar field, $F_{\rho\sigma}=\partial_{\rho}A^{\sigma}-\partial_{\sigma}A^{\rho}$, and $A_{\nu}$ is the electromagnetic four-vector. Moreover, due to the spherical symmetry of the metric, we assume that the evolution of the fields is independent of rotation. Therefore, we can neglect the complex evolutions of the angular variables $\theta$ and $\phi$, and decompose the wave function as follows:
\begin{equation}
\Phi(t,r,\theta,\phi)=\frac{e^{-i\omega t}}{r}\Psi(r)Y_{l}^{m}(\theta,\phi),
\end{equation}
Here, $\omega$ corresponds to the frequency of the wave function, $\Psi$ is the radial wave function, $l$ is the multipole angular quantum number with $l = 0, 1, 2,\cdots$, $m$ corresponds to the magnetic quantum number, and in our discussion, $m$ takes the value of $0$. $Y_{l}^{m}$ is the spherical harmonic function of Hamilton. The transverse equation of equation \eqref{eq:3} is expressed as follows:
\begin{equation}
\frac{1}{\sin\theta}\partial_{\theta}(\sin\theta\partial_{\theta}P_{lm})-\frac{m^{2}}{\sin^{2}\theta}P_{lm}=-l(l + 1)P_{lm},
\end{equation}
To simplify equations \eqref{eq:3} and \eqref{eq:4}, we introduce the tortoise coordinate ($r_{*}$), which is defined as:
\begin{equation}
r_{*}=\int\frac{dr}{f(r)}.
\end{equation}
The introduction of the tortoise coordinate enables the evolution equations of the scalar field and the electromagnetic field to be transformed into a form similar to the Schrödinger equation:
\begin{equation}\label{eq:88}
\frac{d^{2}\psi_{s}}{dr_{*}^{2}}+(\omega^{2}-V(r))\psi_{s}=0.
\end{equation}

\begin{widetext}
\begin{equation}
V(r)=\left(e^{-\frac{2\eta}{r}e^{-\frac{r}{\lambda_G}}}-\frac{2M}{r}\right)(\frac{l(l + 1)}{r^2}+\frac{(1-s)\left(\frac{2M}{r^2} + e^{-\frac{2\eta}{r}e^{-\frac{r}{\lambda_G}}}\left(2 \eta e^{-\frac{r}{\lambda_G}}\frac{1}{r^2}+\frac{2\eta e^{-\frac{r}{\lambda_G}}}{r\lambda_G}\right)\right)}{r}).
\end{equation}
\end{widetext}

Here, $l$ is the multipole angular quantum number, which is restricted by the spin parameter $s$: $l\geq s$. When $s = 0$, it is the effective potential of the scalar field, and when $s = 1$, it corresponds to the effective potential of the electromagnetic field. 

\subsection{Axial Gravitational Perturbations of the Black Hole Model of Dark Matter under the Dominance of Dark Energy}

Perturbations of the gravitational field are important means for studying the structure of black holes, as they can reveal information about the geometry of spacetime \cite{Annulli:2018quj}. In the case of studying the excitation of the spin field, gravitational perturbations are also known as metric perturbations. It consists of the following two parts:
Background metric: It describes the static spacetime where the black hole is located.Linear perturbation: It describes the impact of small perturbations (such as waves, falling particles or celestial bodies) on the background metric. In this case, the metric tensor can be approximately expressed as: 

\begin{equation}\label{eq:10}
g_{\mu\nu} = {^{\scriptscriptstyle(0)}}{g}_{\mu\nu} + h_{\mu\nu}.
\end{equation}
Among them, the background metric of the black hole in the dark matter can be expressed as
\begin{equation}
{^{\scriptscriptstyle(0)}}{g}_{\mu\nu} = \text{diag}(-f(r), \frac{1}{f(r)}, r^2, r^2 \sin^2 \theta).
\end{equation}
Here, \(f(r)\) describes the geometric structure of the black hole in the dark matter. \(h_{\mu\nu}\) is a linear small perturbation with \(h_{\mu\nu}\ll {^{\scriptscriptstyle(0)}}{g}_{\mu\nu}\). It can be further divided into axial perturbations and polar perturbations;
\begin{equation}
h_{\mu\nu} = h_{\mu\nu}^{\text{axial}} + h_{\mu\nu}^{\text{polar}}.
\end{equation} 

In this paper, we only consider the axial gravitational perturbations, which simplifies the calculations.For axial gravitational perturbations, the Regge-Wheeler (RW) gauge can be adopted. Under this gauge, the metric of the axial gravitational perturbations \cite{Regge:1957td} can be written as: 

\vspace{-2pt}
\begin{equation}\label{eq:13}
h_{\mu\nu}^{\text{axial}} = \begin{pmatrix}
0 & 0 & 0 & h_0 \\
0 & 0 & 0 & h_1 \\
0 & 0 & 0 & 0 \\
h_0 & h_1 & 0 & 0
\end{pmatrix} \sin \theta {\partial\theta} P_l (\cos \theta).
\end{equation}
\vspace{3pt}

$P_l (\cos\theta)$ is the Legendre polynomial of order $l$, which describes the angular dependence. For a static and spherically symmetric black hole, the magnetic quantum number $m = 0$, and $h_o$ and $h_1$ are functions of $t$ and $r$.

The perturbed Christoffel symbols can be expressed as
\begin{equation}
\Gamma^{\mu}_{\nu\lambda}= {^{\scriptscriptstyle(0)}}\Gamma^{\mu}_{\nu\lambda}+\delta\Gamma^{\mu}_{\nu\lambda}.
\end{equation}
$\delta \Gamma_{v\lambda}^{\mu}$ is a small correction caused by the perturbation and is given by
\begin{equation}
\delta \Gamma_{v\lambda}^{\mu} = \frac{1}{2}  {^{\scriptscriptstyle(0)}}g^{\mu\rho}
 (h_{\rho\lambda;v} + h_{\rho v;\lambda} - h_{v\lambda;\rho}).
\end{equation}
The perturbation of the Ricci tensor is then expressed as
\begin{equation}
R_{\mu\nu} = {^{\scriptscriptstyle(0)}}{R}_{\mu\nu} + \delta R_{\mu\nu},
\end{equation}
Here, we neglect the influence of dark matter on the perturbation, because compared with the modified background geometry, such a perturbation is negligible \cite{Konoplya:2003ii}. Therefore, the master equation for the axial gravitational perturbation can be expressed as
\begin{equation}\label{eq:17}
\delta R_{\mu\nu}=0.
\end{equation} 
Substituting equations \eqref{eq:10} and \eqref{eq:13} into the master equation \eqref{eq:17} for the axial gravitational perturbation, the non-zero components $\delta R_{24}$ and $\delta R_{34}$ of the gravitational perturbation equation can be expanded as

\vspace{15pt}
\begin{widetext}
\begin{equation}\label{eq:18}
\frac{\partial^{2} {h_1}}{\partial {t^2}} - \frac{\partial}{\partial r} \left( \frac{\partial}{\partial t} h_0 \right) + \frac{2 }{r} \frac{\partial}{\partial t} h_0 - \frac{(f f' + 2 r - f [2l(l+1) - 2 r f'  - 4f ])}{2r^2} h_1 = 0
\end{equation}
\end{widetext}

and
\begin{equation}\label{eq:19}
-\frac{1}{f} \frac{\partial}{\partial t} h_0 + f \frac{\partial}{\partial r} h_1 + f' h_1 = 0.
\end{equation}

The symbols $'$ and $''$ correspond to the first-order derivative and the second-order derivative respectively. Solve for $\frac{\partial}{\partial t}h_0$ in \eqref{eq:19} and substitute it into \eqref{eq:18}.

After simple algebraic operations and the introduction of new variables.
\begin{equation}
h_1 = \frac{r}{f(r)} \psi,
\end{equation}
and
\begin{equation}
dr_* = \frac{dr}{f(r)}.
\end{equation}
After that, the wave equation for the axial gravitational perturbation of the black hole in dark matter can be obtained from equation \eqref{eq:18}.

\begin{equation}
\frac{\partial^2}{\partial t^2} \psi(t,r) - \frac{\partial^2}{\partial r_*^2} \psi(t,r) + V(r)\psi(t,r) = 0,\label{eq:22}
\end{equation}
,
\begin{widetext}
\begin{equation}
V(r)=\left(e^{-\frac{2\eta}{r}e^{-\frac{r}{\lambda_{G}}}}-\frac{2M}{r}\right)\frac{-6M + l(l + 1)r-\frac{6e^{-\frac{(2\eta e^{-\frac{r}{\lambda_{G}}})}{r}-\frac{r}{\lambda_{G}}}\eta(r+\lambda_{G})}{\lambda_{G}}}{r^{3}}.
\end{equation}
\end{widetext}

We briefly discuss the relevant characteristics of the effective potential surrounding the black hole based on the dark matter. The analysis of the effective potential plays a crucial role in studying the quasinormal modes, oscillation frequencies and damping rates of black holes. By analyzing the potential changes caused by various parameters, we can further study the dependence of the QNMs under different parameters.

\begin{figure*}[htbp]
\centering
\begin{subfigure}[b]{0.3\textwidth}
    \includegraphics[width=\textwidth]{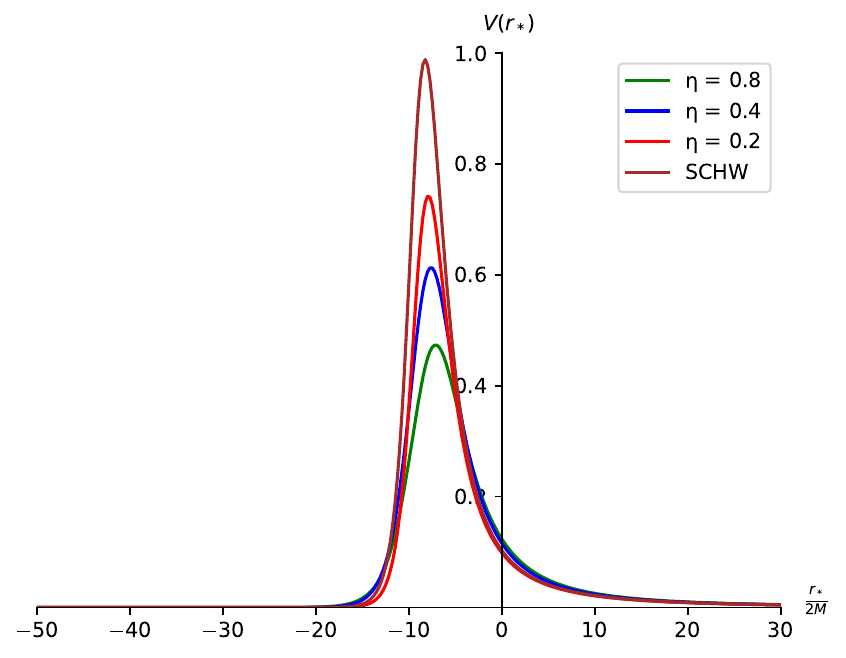}
\end{subfigure}
\hfill
\begin{subfigure}[b]{0.3\textwidth}
    \includegraphics[width=\textwidth]{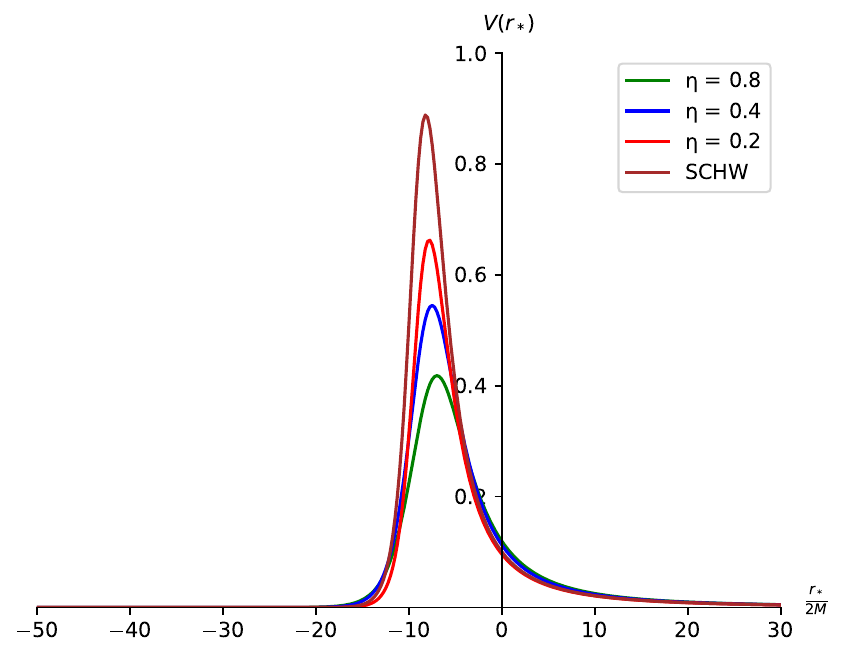}
\end{subfigure}
\hfill
\begin{subfigure}[b]{0.3\textwidth}
    \includegraphics[width=\textwidth]{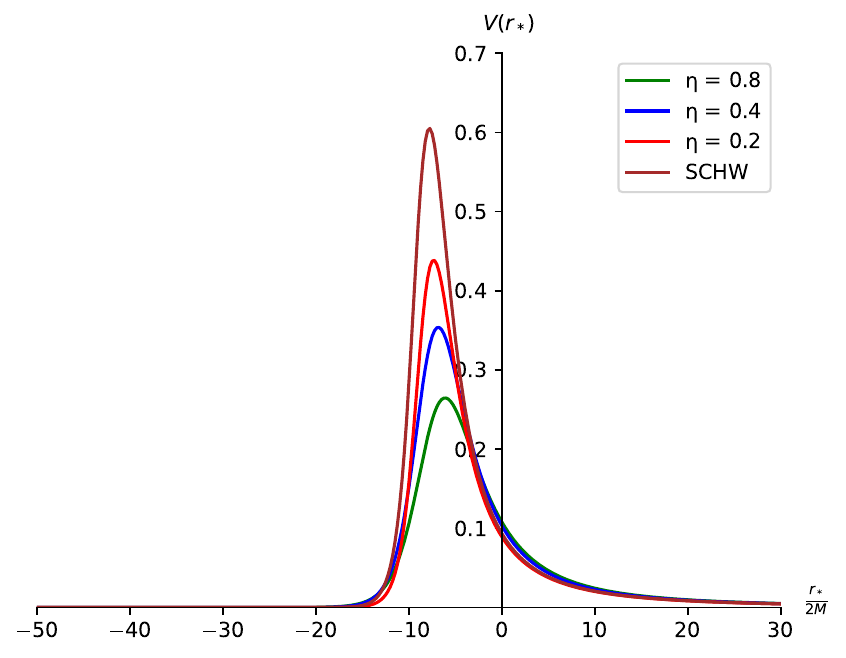}
\end{subfigure}
\captionsetup{justification=raggedright,singlelinecheck=false} 
\caption{It shows how the effective potential energies of the scalar field, electromagnetic field and gravitational field of the black hole surrounded by the dark matter dominated by dark energy vary with the tortoise coordinate under specific parameter settings. Here, we set \(M = 1/2\), \(l = 2\), and \(\lambda_G = 2\), and analyze the changes in the effective potential energy under different values of \(\eta\).}
\label{fig:3}
\end{figure*}

\begin{figure*}[htbp]
\centering
\begin{subfigure}[b]{0.3\textwidth}
    \includegraphics[width=\textwidth]{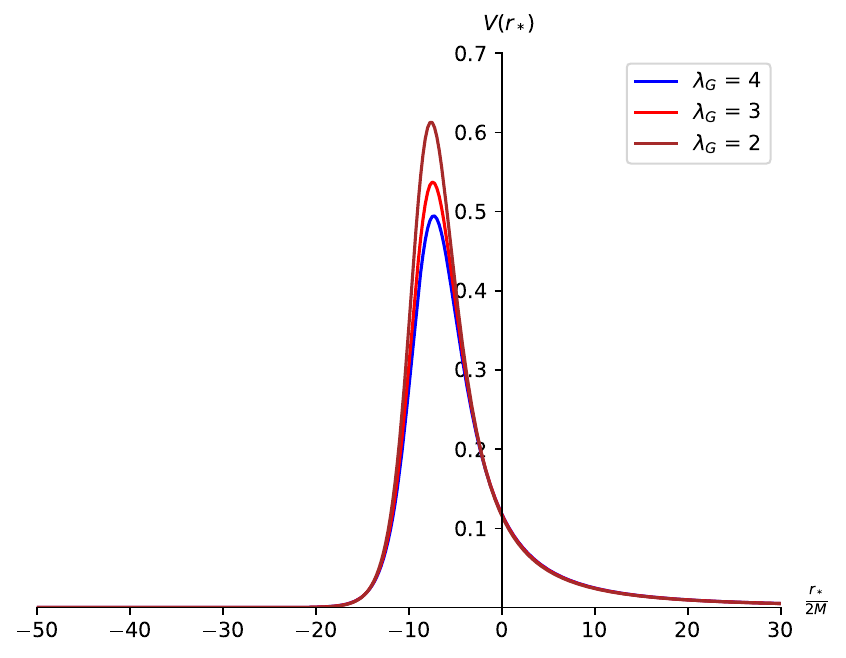}
\end{subfigure}
\hfill
\begin{subfigure}[b]{0.3\textwidth}
    \includegraphics[width=\textwidth]{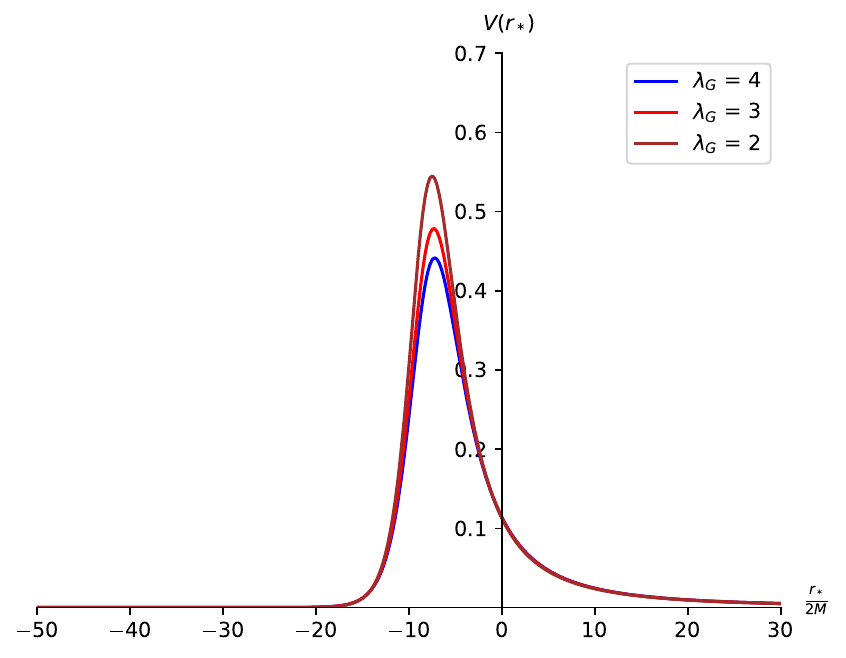}
\end{subfigure}
\hfill
\begin{subfigure}[b]{0.3\textwidth}
    \includegraphics[width=\textwidth]{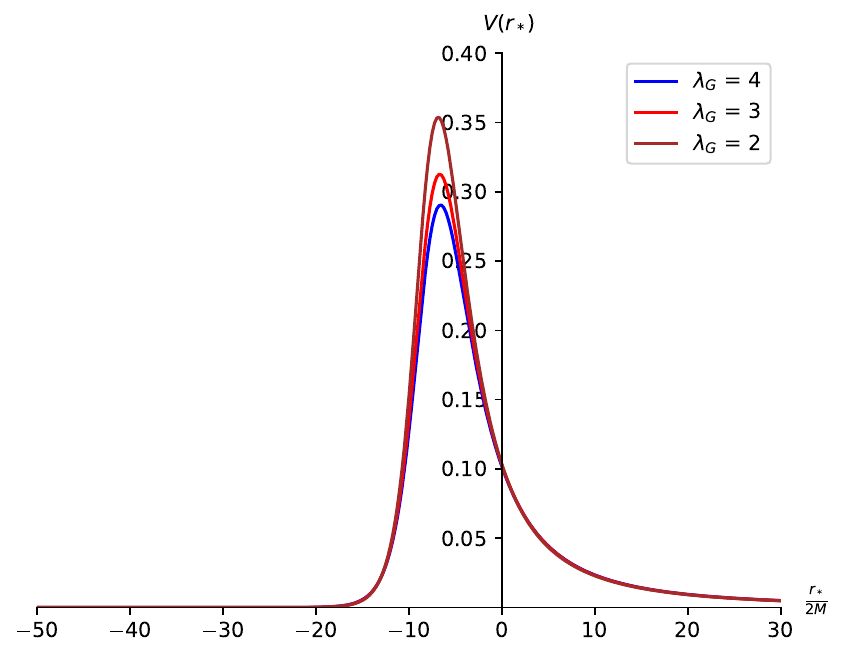}
\end{subfigure}
\captionsetup{justification=raggedright,singlelinecheck=false} 
\caption{It shows how the effective potential energies of the scalar field, electromagnetic field and gravitational field of the black hole surrounded by the dark matter dominated by dark energy vary with the tortoise coordinate. Here, we set \(M = 1/2\), \(l = 2\), and \(\eta = 0.4\), and analyze the changes in the effective potential energy under different values of \(\lambda_G\).}
\label{fig:44}
\end{figure*}

As shown in Figure \ref{fig:3} and Figure \ref{fig:44}, the potential energy curves reveal the changing trends under different gravitational field perturbations. With the increase of the parameter \(\eta\), the potential barrier shows a decreasing trend compared with that of the Schwarzschild black hole (\(\eta = 0\)), which means that the black hole has stronger resistance to perturbations when the value of \(\eta\) is lower. Specifically, a higher potential barrier usually makes it more difficult for perturbations to penetrate, thus resulting in enhanced resistance of the black hole to perturbations. In addition, when studying the dynamics of complex systems, the influence of the spin parameter on the dynamic behavior of the system cannot be ignored. In Figure \ref{fig:3} and Figure \ref{fig:44}, the left, middle and right figures respectively represent the effective potential energy change curves of the scalar field (\(s = 0\)), the electromagnetic field (\(s = 1\)) and the axial gravitational perturbation (\(s = 2\)). The analysis of the dependence of the potential barrier on the spin parameter \(s\) shows that as \(s\) increases, the height of the potential barrier decreases significantly.

In the face of a higher potential barrier, the perturbation waves will encounter greater resistance and the energy consumption speed will also accelerate, which leads to a more rapid change in the spacetime structure near the black hole. Therefore, it is particularly important to study the response degree of perturbations to the dynamic characteristics of black holes, and these changes directly affect the QNMs spectrum of black holes. 

\section{Numerical Methods for QuasiNormal Modes}\label{sec:3}

The ringdown phase of gravitational wave oscillations is a complex nonlinear physical process, and its underlying mechanism can be analyzed in depth through the theory of quasinormal modes \cite{Chandrasekhar:1984siy}. When the spacetime structure of a black hole is perturbed, a series of discrete vibration modes will be excited, and these modes represent the inherent frequencies and damping characteristics of the black hole \cite{Konoplya:2011qq,Isi:2021iql}. In this section, we will use the WKB (Wentzel–Kramers–Brillouin) method and the time-domain method to study the characteristics of these vibration modes. 

\subsection{WKB Method}

Over the present period of time, researchers use various methods to study quasinormal modes, and different methods have their own advantages and limitations. Among them, a particularly elegant method is the WKB (Wentzel-Kramers-Brillouin) approximation method. This method is initially proposed by Schutz and Will \cite{Schutz:1985km} in the context of black hole scattering theory and is subsequently extended by Iyer, Will, Konoplya \cite{Iyer:1986np, Iyer:1986nq, Konoplya:2003ii} and others. When applying the WKB approximation method, we usually focus on the radial part $\psi(r)$ of the perturbation field. This perturbation field has already separated the time dependence through Fourier transformation and the angular dependence through scalar, vector or tensor spherical harmonics suitable for specific problems. Equations \eqref{eq:88} and \eqref{eq:22} can be reformulated as: 

\begin{equation}\label{eq:24}
\frac{d^2 \psi}{d r_*^2} + \left( \omega^2 - V_{\text{eff}} \right) \psi = 0  
\end{equation} 

Here, $r_*$ is the "tortoise coordinate", which is defined in such a way that as the black hole horizon is approached ($r \rightarrow r_p$), $r_* \rightarrow -\infty$, and when approaching spatial infinity ($r \rightarrow \infty$), $r_* \rightarrow \infty$. The underlying potential energy function $V_{\text{eff}}(r_*)$ tends to constant values at the two limits where $r_* \rightarrow \pm \infty$, although these two constants can be different. In particular, $V_{\text{eff}}(r_*)$ reaches its maximum value near $r_* = 0$. The asymptotically flat form of the perturbation field can be described as: 

\begin{equation}  
\psi \sim e^{-i \omega r_*}, \quad r_* \to -\infty 
,
\end{equation}  
\begin{equation}  
\psi \sim e^{i \omega r_*}, \quad r_* \to +\infty . 
\end{equation}

The boundary conditions of the black hole event imply that there are only incoming waves at the event horizon of the black hole (since no waves can escape), while at spatial infinity there are only outgoing waves and no waves enter. The numerical method using the sixth-order WKB formula is employed to determine the QNMs. This method relies on performing WKB expansions on the wave function both at the event horizon and in the far spatial region, and then matching them after the Taylor expansion of the effective potential at the peak point of the potential barrier.  This potential has two extreme points and exhibits the characteristic of monotonic decay. Therefore, the following equation needs to be solved: 

\begin{equation}  
\frac{i (\omega^2 - V_0)}{\sqrt{-2 V_0''}} \bigg|_{r =  \overline{r}_*} - \sum_{i=2}^{6} \tilde{\Lambda}_i = n + \frac{1}{2}.  
\end{equation}

Among them, $V_0$ represents the maximum value of the effective potential, and $V_0''$ is the second-order derivative with respect to the tortoise coordinate. $\Lambda_j (n)$ is the correction term of the $j$-th order WKB method \cite{Ching:1995tj}, and $r =\overline{r}_* $ is the tortoise coordinate at the maximum potential energy of the QNMs. Similarly, in this work, we mainly calculate the frequencies of the fundamental quasinormal modes, that is, the case where $n = 0$. 

\subsection{Time-domain Method}

In order to obtain the dynamic evolution in the spacetime under the gravitational perturbation of the black hole in the M87* galaxy, we adopted the time-domain method. Firstly, the gravitational perturbation wave equation \eqref{eq:24} was reformulated in the form of using the light-cone coordinates $u = t - r_*$ and $v = t + r_*$. Under this transformation, this axial gravitational perturbation equation can be reformulated under the time-domain method as follows:

\vspace{-5pt}
\begin{equation}\label{eq:28}
4\frac{\partial^2 \psi}{\partial u \partial v} + V(r) \psi = 0
\end{equation}
\vspace{-5pt}

Equation \eqref{eq:28} is directly related to the effective potential of the effective dark matter black hole. By utilizing the difference formula, we can discretize this equation, and the relationship between adjacent elements is:

\vspace{5pt}
\begin{widetext}
\begin{equation}\label{eq:29}
\psi(N) = \psi(W) + \psi(E) - \psi(S) - \Delta^2 \frac{(V(W) \psi(W) + V(E) \psi(E))}{8} + o(\Delta^4)
\end{equation} 
\end{widetext}
\vspace{5pt}

Among them, the marking methods for each point are as follows: $N=(u+\Delta,v+\Delta)$, $W=(u+\Delta,v)$, $E=(u,v+\Delta)$, $S=(u,v)$. If we know the values of $\psi$ at points $W$, $E$ and $S$, we can obtain the value of $\psi$ at point $N$ through the difference equation \eqref{eq:29}.

\begin{equation}  \label{eq:30}
\psi(u=u_0, v) = e^{-\frac{(v-v_c)^2}{2\sigma^2}}, \quad \psi(u, v=v_0) = 0  
\end{equation}  
The initial data are set to satisfy equation \eqref{eq:30} on the null surfaces $u = u_0$ and $v = v_0$. Here, $v_c$ is a parameter of the Gaussian wave packet.

In order to extract the frequencies of the quasinormal modes from the obtained time-domain profiles, we adopted the Prony method, which fits the signal by representing it as a sum of damped exponential functions:

\begin{equation}  
\psi(t) \approx \sum_{i=1}^{p} C_i e^{-i \omega_i t}  
\end{equation} 

\section{Numerical Results of QuasiNormal Modes of Black Holes in the Dark Matter of M87*}
\label{sec:4}

\begin{figure*}[tbp]
\centering
\begin{subfigure}[b]{0.3\textwidth}
    \includegraphics[width=\textwidth]{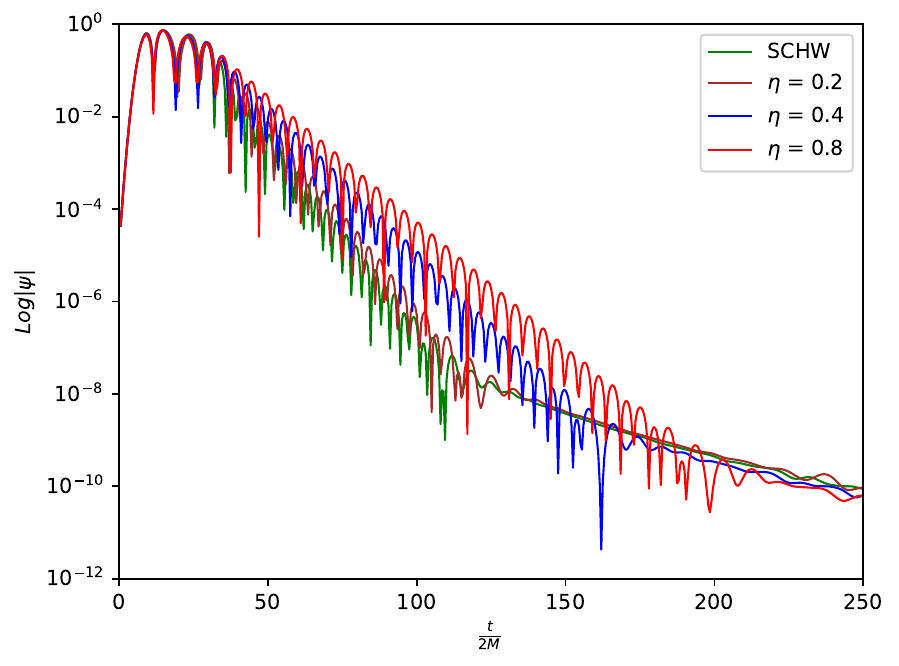}
\end{subfigure}
\hfill
\begin{subfigure}[b]{0.3\textwidth}
    \includegraphics[width=\textwidth]{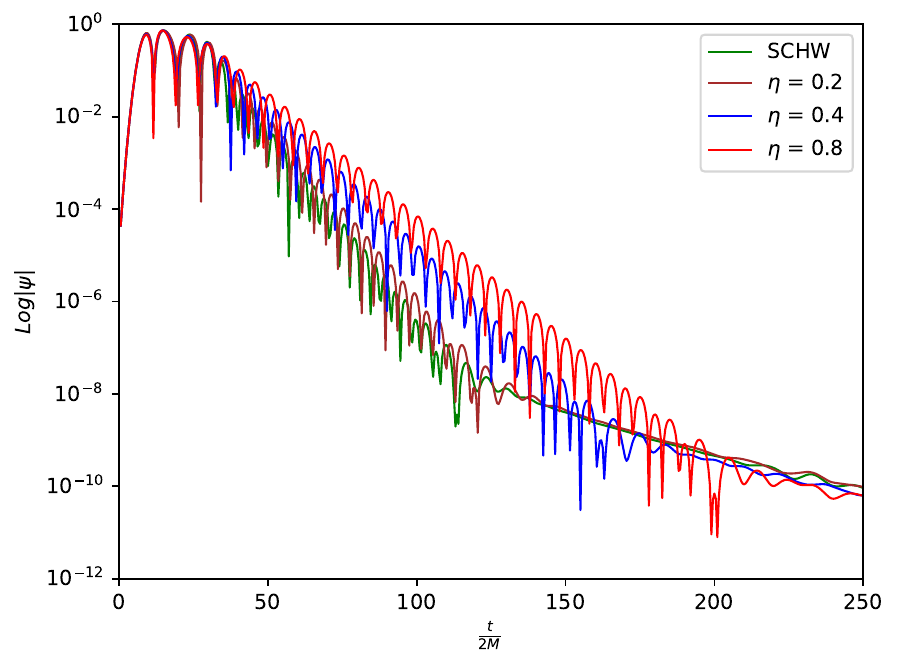}
\end{subfigure}
\hfill
\begin{subfigure}[b]{0.3\textwidth}
    \includegraphics[width=\textwidth]{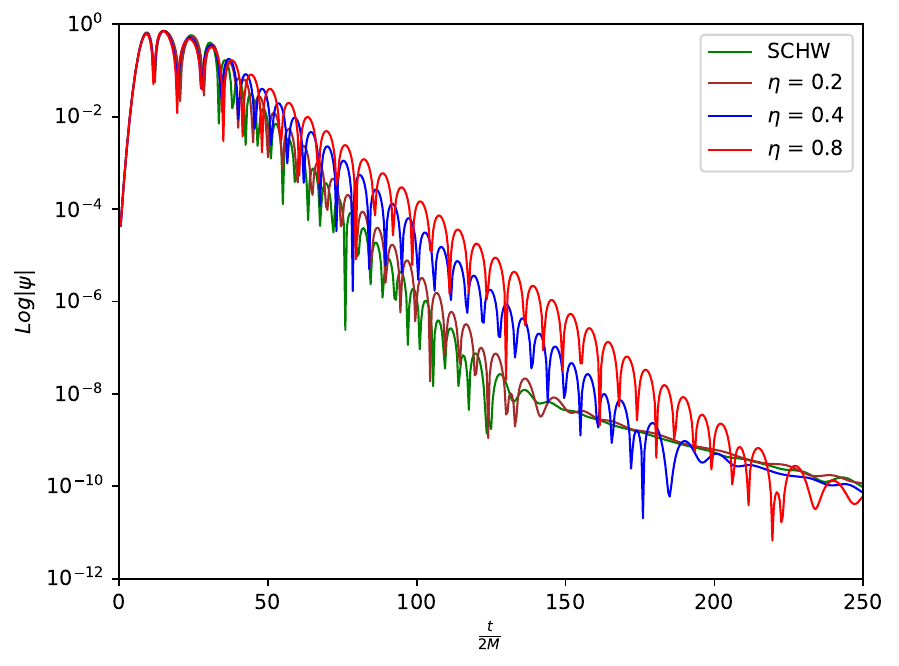}
\end{subfigure}
\captionsetup{justification=raggedright,singlelinecheck=false} 
\caption{It shows the comparison of the perturbation time-domain profiles (for the massless scalar field, electromagnetic field and gravitational field) under different values of $\eta$ with the fixed parameters  $M = 1/2$, $l = 2$, and $\lambda_G = 2$.}
\label{fig:4}
\end{figure*}

\begin{figure*}[btp]
\centering
\begin{subfigure}[b]{0.3\textwidth}
    \includegraphics[width=\textwidth]{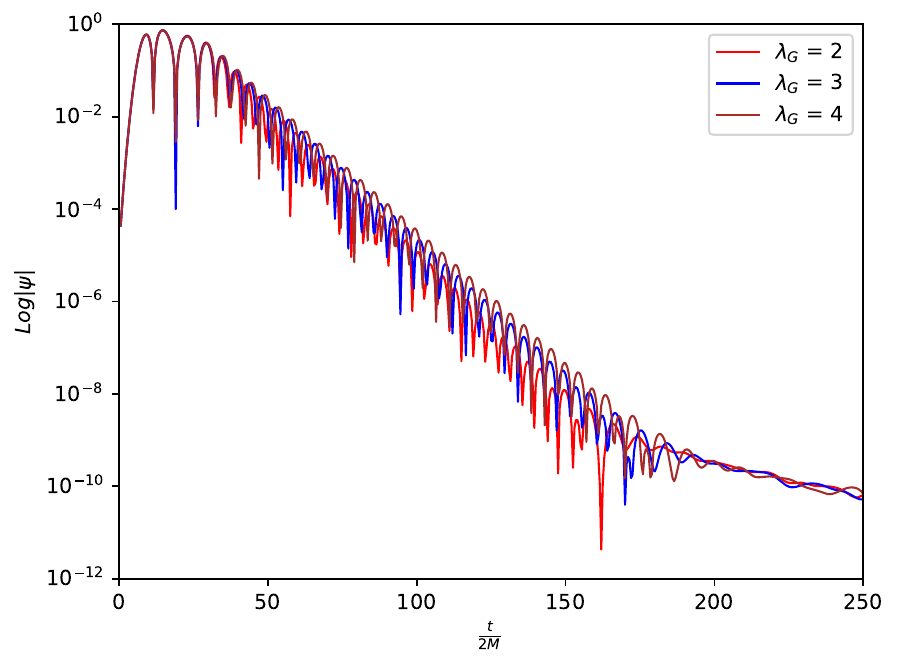}
\end{subfigure}
\hfill
\begin{subfigure}[b]{0.3\textwidth}
    \includegraphics[width=\textwidth]{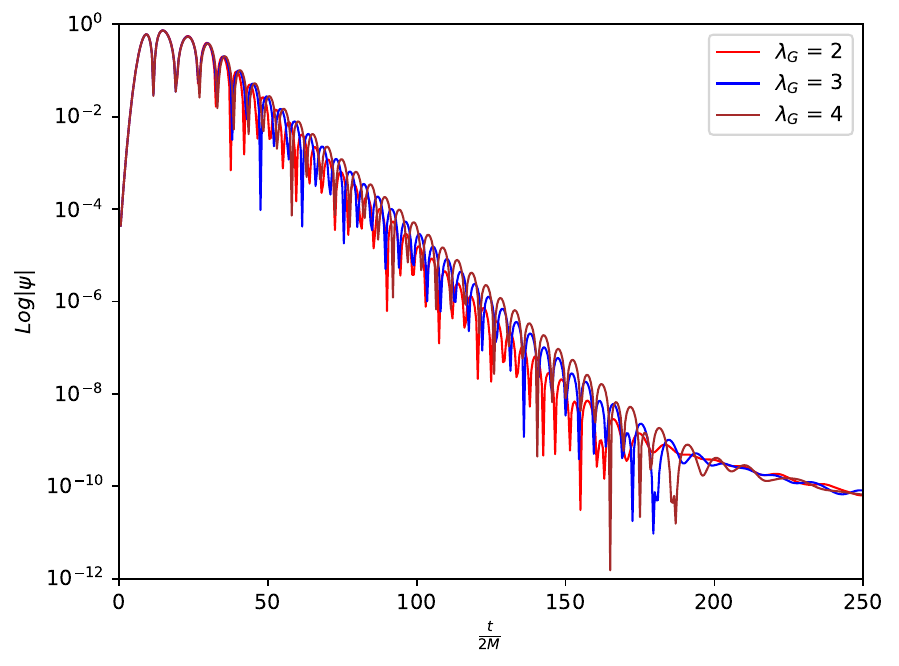}
\end{subfigure}
\hfill
\begin{subfigure}[b]{0.3\textwidth}
    \includegraphics[width=\textwidth]{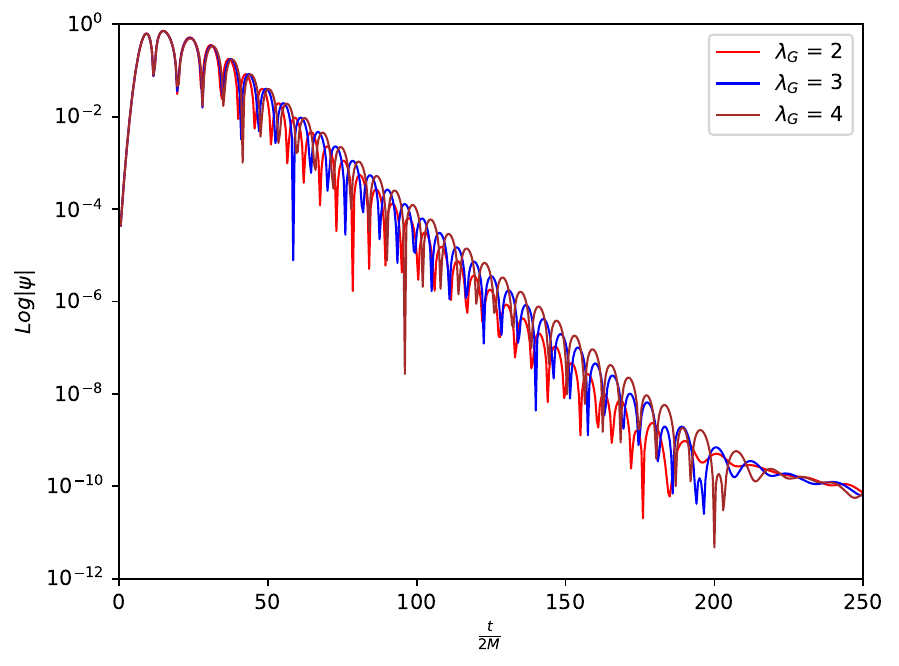}
\end{subfigure}
\captionsetup{justification=raggedright,singlelinecheck=false} 
\caption{It shows the comparison diagrams of the time-domain profiles (for the massless scalar field, electromagnetic field and gravitational field) under different values of $\lambda_G$ with the fixed parameters $M = 1/2$, $l = 2$, and $\eta = 0.4$.}
\label{fig:5}
\end{figure*}

In the previous section, we preliminarily explore the calculation methods for quasinormal modes, with a particular focus on the time-domain method and the WKB approximation. By solving equation \eqref{eq:28}, we obtain the ringing waveforms for different values of the parameters $\lambda_G$ and $\eta$. The evolution of the Gaussian pulse is shown in Figure \ref{fig:4} and Figure \ref{fig:5}. 

The left, middle and right figures respectively correspond to the quasinormal mode curves of the scalar field, electromagnetic field and axial gravitational field perturbations. In Figure \ref{fig:4}, we display the time evolution of the quasinormal modes of the black hole at the center of the M87* galaxy in the dark matter and in the vacuum case (i.e., $\eta = 0$). In the case of $\eta = 0$, the system is simplified to the classic Schwarzschild black hole model, and the attenuation characteristics of the wave packet are consistent with the predictions of the classical gravitational field theory. It can be seen that as the parameter $\eta$ increases, the influence of $\eta$ on the perturbation waveform is not obvious in the initial stage of oscillation. However, when the system enters the quasinormal mode stage, the increase in the parameter causes the oscillation waveform to become flatter compared to the case of the Schwarzschild black hole, and the attenuation time of the wave packet also increases significantly. This indicates that the damping effect of the system is weakened, meaning that the system has a more persistent response to external excitations. 

We used the Prony method to extract the first-order (n = 0) quasinormal mode frequencies from Figure \ref{fig:4} and Figure \ref{fig:5}. This method fits N data points by selecting the required number of pure damped exponential functions. When calculating the quasinormal mode of scalar perturbations for the Schwarzschild case (l = 2), the result we obtained was \(0.967771 - 0.192623i\). This result shows a good agreement compared with the result \(0.967284 - 0.193532i\) obtained by the 6-th WKB method and the result \(0.967284 - 0.193532i\) obtained by applying the continued fraction method in \cite{Daghigh:2020mog}. This indicates that the effectiveness of the Prony method in extracting quasinormal mode frequencies has been verified. 

\vspace{5pt}
\begin{table}[H]
\centering
\begin{tabular}{ccccc}
\hline
\multicolumn{1}{c}{Parameter $\eta$} & \multicolumn{1}{c}{WKB} & \multicolumn{1}{c}{Prony} \\
\hline
0 & 0.967284 - 0.193532i & 0.967771 - 0.192623i \\
0.2 & 0.836324 - 0.174832i & 0.835016 - 0.175113i \\
0.4 & 0.758817 - 0.162572i & 0.757311 - 0.169861i \\
0.6 & 0.705404 - 0.153791i & 0.702935 - 0.155294i \\
0.8 & 0.665469 - 0.147111i & 0.672412 - 0.144323i \\
\hline
\end{tabular}
\vspace{0pt} 
\caption{Shows the scalar field quasinormal perturbation frequencies of the black hole model measured by the time-domain method and the sixth-order WKB approximation method respectively, under the fixed angular quantum number \(l = 2\) and \(\lambda_G = 2\).} 
\label{table:2}
\end{table}

\vspace{-0pt}
\begin{table}[H]
\centering
\hypertarget{table:3}{}
\begin{tabular}{ccc}
\hline
\multicolumn{1}{c}{Parameter $\lambda_G$} & \multicolumn{1}{c}{WKB} & \multicolumn{1}{c}{Prony} \\
\hline
1 & 0.852855 - 0.188302i & 0.851145 - 0.187735i \\
2 & 0.758817 - 0.162572i & 0.755927 - 0.167602i \\
3 & 0.711526 - 0.148718i & 0.713881 - 0.147522i \\
4 & 0.683449 - 0.140516i & 0.682364 - 0.145589i \\
5 & 0.664891 - 0.135151i & 0.664113 - 0.139778i \\
\hline
\end{tabular}
\caption{Shows the scalar field quasinormal perturbation frequencies of the black hole model measured by the time-domain method and the sixth-order WKB approximation method respectively, under the fixed angular quantum number \(l = 2\) and \(\eta = 0.4\).}
\label{table:3} 
\end{table}

\vspace{-0pt}
\begin{table}[H]
\centering
\begin{tabular}{ccccc}
\hline
\multicolumn{1}{c}{Parameter $\eta$} & \multicolumn{1}{c}{WKB} & \multicolumn{1}{c}{Prony} \\
\hline
0 & 0.915187 - 0.190022\(i\) & 0.917509 - 0.188484\(i\) \\
0.2 & 0.788183 - 0.171141\(i\) & 0.784779 - 0.177274\(i\) \\
0.4 & 0.713348 - 0.158782\(i\) & 0.710422 - 0.159327\(i\) \\
0.6 & 0.661887 - 0.149924\(i\) & 0.669817 - 0.130178\(i\) \\
0.8 & 0.623464 - 0.143176\(i\) & 0.621413 - 0.142861\(i\) \\
\hline
\end{tabular}
\par\medskip
\caption{Shows the electromagnetic field quasinormal perturbation frequencies of the black hole model measured by the time-domain method and the sixth-order WKB approximation method respectively, under the fixed angular quantum number \(l = 2\) and \(\lambda_G = 2\). }
\label{table:4}
\end{table} 

\begin{table}[t]
\centering 

\begin{tabular}{ccccc}
\hline       
\multicolumn{1}{c}{Parameter $\lambda_G$} & \multicolumn{1}{c}{WKB} & \multicolumn{1}{c}{Prony} \\
\hline
1 & 0.801481 - 0.183617\(i\) & 0.799729 - 0.192438\(i\) \\
2 & 0.713348 - 0.158782\(i\) & 0.710426 - 0.159329\(i\) \\
3 & 0.669761 - 0.145518\(i\) & 0.668737 - 0.145744\(i\) \\
4 & 0.644038 - 0.137667\(i\) & 0.643586 - 0.138019\(i\) \\
5 & 0.627088 - 0.132511\(i\) & 0.627706 - 0.131945\(i\) \\
\hline
\end{tabular}
\par\medskip
\caption{\Centering Shows the electromagnetic field quasinormal perturbation frequencies of the black hole model measured by the time-domain method and the sixth-order WKB approximation method respectively, under the fixed angular quantum number \(l = 2\) and \(\eta = 0.4\).}
\label{table:5}
\end{table}

\begin{table}[t]
    \centering 
    \begin{tabular}{c c c} 
        \hline
       \multicolumn{1}{c}{Parameter $\eta$} & \multicolumn{1}{c}{WKB} & \multicolumn{1}{c}{Prony} \\
        \hline
        0   & 0.747239 - 0.177782\(i\) & 0.743269 - 0.178926\(i\) \\
        0.2 & 0.633375 - 0.158329\(i\) & 0.637437 - 0.158333\(i\) \\
        0.4 & 0.567531 - 0.145531\(i\) & 0.567244 - 0.148855\(i\) \\
        0.6 & 0.522754 - 0.136383\(i\) & 0.522974 - 0.139747\(i\) \\
        0.8 & 0.489537 - 0.129549\(i\) & 0.494138 - 0.134863\(i\) \\
        \hline
    \end{tabular}
    \par\medskip
    \captionsetup{justification=raggedright,singlelinecheck=false} 
    \caption{\Centering Shows the axial gravitational field quasinormal perturbation frequencies of the black hole model measured by the time-domain method and the sixth-order WKB approximation method respectively, under the fixed angular quantum number \(l = 2\) and \(\lambda_G = 2\).
}
\label{table:6}
\end{table}

\begin{table}[t]
\centering
\begin{tabular}{ccc}
\hline
\multicolumn{1}{c}{Parameter $\lambda_G$} & \multicolumn{1}{c}{WKB} & \multicolumn{1}{c}{Prony} \\
\hline
1 & 0.638092 - 0.169684\(i\) & 0.636274 - 0.171786\(i\)
 \\
2 & 0.567531 - 0.145531\(i\) & 0.572076 - 0.148765\(i\)
 \\
3 & 0.534968 - 0.133727\(i\) & 0.533639 - 0.138337\(i\)
\\
4 & 0.516443 - 0.126949\(i\) & 0.518367 - 0.129734\(i\) 	
\\
5 & 0.504505 - 0.122549\(i\) & 0.517701 - 0.122213\(i\) 	
\\
\hline
\end{tabular}
\par\medskip
\captionsetup{justification=raggedright,singlelinecheck=false} 
\caption{Shows the axial gravitational field quasinormal perturbation frequencies of the black hole model measured by the time-domain method and the sixth-order WKB approximation method respectively, under the fixed angular quantum number \(l = 2\) and \(\eta = 0.4\). }
\label{table:7}
\end{table}

Tables \ref{table:2} to \ref{table:7} demonstrate the influence of parameters $\eta$ and $\lambda_G$ on the quasinormal mode frequencies of the black hole at the center of the M87* galaxy under different gravitational fields. These frequencies are calculated by the Prony and WKB methods.

Specifically, in Tables \ref{table:2}, \ref{table:4}, and \ref{table:6}, the variations of the quasinormal mode frequencies under vacuum conditions ($\eta = 0$) and under the influence of the dark matter are shown respectively. The results indicate that the real part of the quasinormal mode frequency decreases as the parameter $\eta$ increases, while the imaginary part is negative and its absolute value decreases significantly, which is clearly different from the performance of the Schwarzschild black hole. The previous analyses of the effective potential and the quasinormal mode curves provide preliminary evidence for this conclusion.In Table \ref{table:2}, when $\eta = 0.2$, the quasinormal frequency is $0.836324 - 0.174832i$. In contrast, for the Schwarzschild black hole ($\eta = 0$) with a quasinormal frequency of $0.967284 - 0.193532i$, the real part of the frequency is reduced by around $13.5\%$, and the absolute value of the imaginary part is decreased by about $9.7\%$. This shows that there are distinct oscillation and damping characteristics. 

Then, Tables \ref{table:3}, \ref{table:5}, and \ref{table:7} also present the quasinormal mode frequencies of the black hole at the center of the M87* galaxy in the dark matter under the different field perturbations. It can be observed that as $\lambda_G$ increases, the effect of gravity is constrained on a large scale. This phenomenon is reflected in the decreasing absolute values of both the real and imaginary parts of the QNMs frequencies, indicating a significant weakening of the oscillatory characteristics of the system, manifested as the prolongation of the oscillation period and the reduction of the response speed, accompanied by the decrease in the damping rate, resulting in the persistence of the oscillation state.

\section{Summary}

In this study, we explore the spacetime characteristics of black holes by studying the quasinormal modes in the dark matter induced by dark energy. Employing the Prony method and the WKB approximation, our results show that the dark matter parameter $\eta$ and the Newtonian gravitational shielding parameter $\lambda_{G}$ have a significant impact on the quasinormal mode frequencies within the considered parameter range. Specifically, as $\eta$ and $\lambda_{G}$ increase, the real part of the quasinormal mode frequencies gradually decreases, and the absolute value of the imaginary part also shows a decreasing trend. This indicates that the existence of dark matter and the Newtonian gravitational shielding effect will lead to a reduction in the oscillation frequencies of the black hole spacetime and a slowdown in the energy decay rate. Without loss of generality, we take $\lambda_{G}=2$ as an example, when the dark matter parameter $\eta$ increases from $0$ to $0.2$, the real part of the quasinormal mode frequencies decreases by approximately $13.5\%$, and the absolute value of the imaginary part decreases by approximately $9.7\%$ (see Table II). These results clearly demonstrate that the influence of dark matter on the oscillation and energy decay of the black hole spacetime cannot be ignored. Compared with black holes in a vacuum, the existence of the dark matter significantly changes the quasinormal mode frequencies of black holes. Further analysis shows that within the studied parameter space ($\lambda_{G}$ and $\eta$), the imaginary part of the quasinormal mode frequencies is negative under all perturbation modes, indicating that the considered black hole model maintains linear dynamical stability in this dark matter environment.

\section*{Acknowledgments}
This research partly supported by the
National Natural Science Foundation of China (Grant No. 12265007).

\newpage



\bibliography{ref}
\bibliographystyle{apsrev4-1}

\end{document}